\documentclass[12pt]{iopart}
\usepackage{epsfig,amssymb}

\newcommand{\be}{\begin{equation}}
\newcommand{\ee}{\end{equation}}
\newcommand{\beq}{\begin{eqnarray}}
\newcommand{\eeq}{\end{eqnarray}}
\newcommand{\sss}{\scriptscriptstyle}

\def\lsim{\hbox{ \raise.35ex\rlap{$<$}\lower.6ex\hbox{$\sim$}\ }}
\def\gsim{\hbox{ \raise.35ex\rlap{$>$}\lower.6ex\hbox{$\sim$}\ }}

\begin{document}
\title{Cusps on cosmic superstrings with junctions}
\author{Anne-Christine Davis$^1$,
  William Nelson$^2$,\\ Senthooran
  Rajamanoharan$^1$, Mairi
  Sakellariadou$^2$}
\address{$^1$DAMTP, Centre for Mathematical Sciences, Wilberforce
  Road, Cambridge CB3 0WA, U.K.}  \address{$^2$Department of Physics,
  King's College, University of London, Strand WC2R 2LS, London, U.K.}
\eads{\mailto{a.c.davis@damtp.cam.ac.uk},\;\mailto{william.nelson@kcl.ac.uk},
  \;\mailto{s.rajamanoharan@damtp.cam.ac.uk}
  ,\;\mailto{mairi.sakellariadou@kcl.ac.uk}}

\begin{abstract}
 The existence of cusps on non-periodic strings ending on D-branes is
demonstrated and the conditions, for which such cusps are generic, are
derived. The dynamics of F-, D-string and FD-string junctions are
investigated. It is shown that pairs of FD-string junctions, such as
would form after intercommutations of F- and D-strings, generically
contain cusps.  This new feature of cosmic superstrings opens up the
possibility of extra channels of energy loss from a string network.
The phenomenology of cusps on such cosmic superstring
networks is compared to that of cusps formed on networks of their
field theory analogues, the standard cosmic strings.

\vspace{.2cm}
\noindent
\end{abstract}
 

%\maketitle

%%%%%%%%%%%%%%%%%%%%%%%%%%%%%%%%%%%%%%%%%%%%%%%%%%%%%%%%%%%%%%%%%%%%%%%%%
\section{Introduction}

Fundamental (F) strings and Dirichlet branes with one non-compact
spatial dimension (D-strings) are generically
formed~\cite{sarangi-tye,ma,jonesetal,dvalietal} at the end of brane
inflation~\cite{dvali-tye99} within the context of string inspired
cosmological models.  Such strings, known as cosmic superstrings, are
of cosmological size and could play the r\^ole of cosmic
strings~\cite{Vilenkin_shellard,ms-cs07}, false vacuum remnants formed
generically at the end of hybrid inflation within Grand Unified
Theories~\cite{jrs03,ms-cs08}. Cosmic superstrings have gained a lot
of interest, particularly since it is believed that they may be
observed in the sky, providing both a means of testing string theory
and a hint for a physically motivated inflationary model (for a recent
reviews, see e.g. Ref.~\cite{sakellariadou2008,Davis2005,Davis2008}).
 
The most significant difference between cosmic superstrings and the
field theory cosmic strings we are more familiar with, is the
existence of three string junctions, the presence of which could
strongly effect the dynamics of the string network.  Understanding
these new dynamical effects is critical if there is to be any hope of
differentiating cosmic superstrings from their solitonic analogues. A
number of
analytical~\cite{Copeland:2006if,Copeland:2006eh,Copeland:2007nv} and
numerical~\cite{Sakellariadou:2004wq,Avgoustidis:2004zt,Copeland:2005cy,Hindmarsh:2006qn,Rajantie:2007hp,Urrestilla:2007yw,Sakellariadou:2008ay}
studies have addressed cosmic superstring dynamics. We note that, in
principle, cosmic superstring dynamics ought to be studied using the
Dirac-Born-Infeld action, the low-energy effective action for many
varieties of strings arising in the context of string theory.

In what follows, we investigate (generic) string solutions ending on
parallel Dirichlet branes as a pedagogical example of the effects
possible at a three string junction. We then look at the dynamics of a
junction made up of an F-string, a D-string, and an FD-string, which
are expected to form through intercommuting of initial configurations
composed from F- and D-string networks.  We show that cusps are
generic features for such strings, opening up a new energy loss
mechanism for the network, in addition to the formation and subsequent
decay of closed loops and the formation of bound
states~\cite{Sakellariadou:2008ay}. Studies of the phenomenological
implications of cosmic superstrings, particularly the
gravitational~\cite{Damour:2004kw,Siemens:2006vk,Siemens:2006yp,Urrestilla:2007yw,hogan}
and Ramond-Ramond~\cite{Sakellariadou:2004wq,Firouzjahi:2007dp}
radiation emitted from cosmic superstrings --- predominantly from
cusps and to some extent from kinks --- are then justified. Some of
the phenomenological consequences of cosmic superstring dynamics are
discussed here.

%%%%%%%%%%%%%%%%%%%%%%%%%%%%%%%%%%%%%%%%%%%%%%%%%%%%%%%%%%%%%%%%%%%%%%%%%%%
\section{DBI string ending on parallel D-branes}\label{sec:Dbranes}

The world-history of a string can be represented by its world-sheet
$$x^\mu=x^\mu(\tau,\sigma);$$ a
two-dimensional surface in the four-dimensional space-time.  The
world-sheet's coordinates $\tau, \sigma$ are arbitrary time-like
and space-like parameters, respectively.  A metric for the
two-dimensional world-sheet is induced by pulling-back the space-time
metric
$$\gamma_{\alpha\beta}=g_{\mu\nu} x^\mu_{,\alpha}x^\nu_{,\beta}~,$$
where $g_{\mu\nu}$ denotes the four-dimensional metric.

Consider a Dirac-Born-Infeld (DBI) string in a Minkowski background,
with end-points that are constrained to lie on two stationary,
parallel and flat D$n$-branes, where $n$ is the spatial dimensionality
of the branes. Without loss of generality, we choose Cartesian
space-time coordinates in which the separation vector between the two
branes lies in the $z$-direction.

The action for the DBI string is 
\be S=-\mu\int{\rm d}\tau{\rm
  d}\sigma\sqrt{-|\gamma_{\alpha\beta}+\lambda F_{\alpha\beta}|}\ ,
\ee 
where $\mu$ is the string tension parameter and $\lambda=2\pi\alpha'$.
For a $(p,q)$-string (a bound state of $p$ coincident F-strings and
$q$ coincident D-strings) the string tension is $\mu=|q|/(g_s
\lambda)$, where $g_s$ is the perturbative string coupling.
The electromagnetic field strength, $F_{\alpha\beta}$, associated with
the $U(1)$ gauge field $A_\alpha$ on the world-sheet, reads
$$F_{\alpha\beta} = \partial_\alpha A_\beta - \partial_\beta A_\alpha~.$$ The
DBI action describes the low-energy dynamics of certain classical
string-like objects found in string theory.

Due to the re-parametrisation invariance of the string world-sheet
$x^{\mu}(\tau,\sigma)$, we are free to impose the conformal gauge condition:
\be
 \dot{x}^2 + x'^2 = 0~\quad\mathrm{and}\quad\dot{x}\cdot x'=0~,
\ee
where $\dot{}\equiv \partial/\partial{\tau}$ and $'\equiv
\partial/\partial{\sigma}$.  In this gauge, the DBI equation of motion
for the string world-sheet reads~\cite{Copeland:2007nv}
\be
 \ddot{x} - {x}'' = 0~.
\ee
The electric flux along the string,
\be
p\equiv \frac{\partial \cal{L}}{\partial F_{\tau\sigma}} =\frac{\lambda^2\mu
F_{\tau\sigma}}{\sqrt{-x'^{2}\dot{x}^{2}-\lambda^2{F_{\tau\sigma}}^2}}
\ ,
\ee
is a conserved quantity; in the $(p,q)$-string picture it corresponds
to the number of coincident F-strings that make up the bound state.

The conformal gauge condition allows for a residual gauge symmetry,
which can be fixed in a manner that is consistent with the equations
of motion by imposing the temporal gauge condition $x^0 \equiv t =
\tau$. Once we have totally fixed the world-sheet parametrisation in
this way, the equation of motion reads
\be
 \ddot{\bf x} - {\bf x}'' = 0~,
\label{eom}
\ee
with gauge constraints
\be\label{eq:gauge_con}
 \dot{\bf x}^2 + {\bf x}'^2 = 1\quad\mathrm{and}\quad
 \dot{\bf x}\cdot{\bf x}'=0\ ,
\ee
where $x^{\mu}(t,\sigma)=(t,{\bf x}(t,\sigma))$.

The general solution of the equation of motion, Eq.~(\ref{eom}), is
\be
 {\bf x} = \frac{1}{2} \left[ {\bf a} ( t-\sigma ) + {\bf b}
( t+\sigma) \right]~,
\ee
where ${\bf a}( t-\sigma )$ and ${\bf b}( t+\sigma )$ are arbitrary
vector-valued functions. However, the gauge constraints,
Eq.~(\ref{eq:gauge_con}), impose the additional (necessary and
sufficient) restrictions
\be\label{eq:gauge_con2}
 |{\bf a}'|^2 = |{\bf b}'|^2 = 1~,
\ee
where primes ($^{'}$) here refer to total derivative of these
single-variable functions. Therefore, prior to taking boundary
conditions into account, a DBI string solution may be completely
specified by the arbitrary choice of two parametrised curves on the
unit sphere.

Let us consider what additional conditions are imposed by the boundary
conditions for the current problem -- that of a DBI string ending on
two stationary and parallel D$n$-branes. Without loss of generality,
we choose the parametrisation of the world-sheet so that at time $t$
the end-point of the string attached to the upper brane (towards the
$+z$-direction) is at $\sigma=0$, whilst the end-point of the string
attached to the lower brane is at $\sigma=L(t)$. Note that $L(t)$
denotes the \emph{parameter length} of the string (as opposed to its
physical length).  When the string has time-independent boundary
conditions (as is the case considered here), the parameter length,
$L(t)$, is assumed to be time-independent.  However, for more general
configurations (which we consider later on), it is important to look
for solutions where $L(t)$ is not necessarily constant. Thus, we do
not make here the (usual) assumption that $\dot{L}(t)=0$. Certainly,
when we analyse a time-independent brane configuration, we recover the
expected result, namely that the parameter length of the string
remains constant.

It turns out to be convenient to decompose all spatial three-vectors
(e.g., ${\bf v}$) into a part parallel to the D-branes and one
perpendicular to them:
\be
{\bf v} =  {\bf v}_{\parallel} + {\bf v}_{\perp}\ .
\ee
This decomposition (which is clearly unique), allows us to distinguish
between the boundary conditions for Neumann directions (parallel to
the branes) and the boundary conditions for Dirichlet directions
(perpendicular to the branes).

On the one hand, the fact that the end-points of the string are
constrained to the two D-branes implies, in Dirichlet directions, that
\be
 \dot{\bf x}_{\perp}(t,0)=0\quad\mathrm{and}\quad
 \dot{\bf x}_{\perp}(t,L(t))+\dot{L}(t){\bf x}'_{\perp}(t,L(t))=0\ ,
\ee
whilst in Neumann directions, that
\be
 {\bf x}'_{\parallel}(t,0)=0\quad\mathrm{and}\quad
 {\bf x}'_{\parallel}(t,L(t))=0\ .
\ee
If we consider what these boundary conditions mean in terms of ${\bf
  a}$ and ${\bf b}$, we find that the boundary conditions at the
$\sigma=0$ end of the string lead to
\beq\label{eq:Dbrane_boun1a}
  {\bf a}'_{\parallel}(t) &=& {\bf b}'_{\parallel}(t)~, \\
  {\bf a}'_{\perp}(t) &=& - {\bf b}'_{\perp}(t)~.\label{eq:Dbrane_boun1b}
\eeq
Geometrically speaking, Eqs.~(\ref{eq:Dbrane_boun1a}),
(\ref{eq:Dbrane_boun1b}) imply that the curves ${\bf a'}(t)$ and
${\bf b'}(t)$ are related by inversion through a surface of identical
dimension and orientation to the D-branes, that passes through the
centre of the unit sphere. On the other hand, the boundary conditions
at the $\sigma=L(t)$ end of the string imply that
\beq\label{eq:Dbrane_boun2a} {\bf a}'_{\parallel}(t-L(t)) &=& {\bf
  b}'_{\parallel}(t+L(t))~, \\ \left[\dot{L}(t)-1\right]{\bf
  a}'_{\perp}(t-L(t)) &=& -\left[\dot{L}(t)+1\right]{\bf
  b}'_{\perp}(t+L(t))~.
  \label{eq:Dbrane_boun2b}
\eeq
Any curves ${\bf a}'$ and ${\bf b}'$ that satisfy
Eqs.~(\ref{eq:Dbrane_boun1a}), (\ref{eq:Dbrane_boun1b}),
(\ref{eq:Dbrane_boun2a}), (\ref{eq:Dbrane_boun2b}) must also
necessarily satisfy the conditions that we obtain by squaring them and
imposing the gauge constraints Eq.~(\ref{eq:gauge_con2}), written in
the form
\be
 |{\bf a}'_{\parallel}|^{2}+|{\bf a}'_{\perp}|^2 =
 |{\bf b}'_{\parallel}|^{2}+|{\bf b}'_{\perp}|^2 = 1\ .
\ee
It is easily checked that the necessary conditions that we have
obtained in this manner can 
only be satisfied when $\dot{L}=0$. We have thus confirmed explicitly
that the string solutions for this stationary brane configuration must
have constant parameter length, $L(t)=L$.

Using $\dot{L}=0$, as well as the boundary conditions
Eqs.~(\ref{eq:Dbrane_boun1a}), (\ref{eq:Dbrane_boun1b}), it is
straightforward to re-write Eqs.~(\ref{eq:Dbrane_boun2a}),
(\ref{eq:Dbrane_boun2b}) in the form 
\be\label{eq:periodic} 
{\bf a}'(t-L)={\bf a}'(t+L)\ , 
\ee 
and similarly for ${\bf b}'$. Geometrically speaking, ${\bf a}'(t)$
and ${\bf b}'(t)$ describe closed curves of parameter length $2L$ on
the unit sphere.

Equations (\ref{eq:Dbrane_boun2a}) and (\ref{eq:Dbrane_boun2b}) are
necessary but not sufficient to satisfy the Dirichlet boundary
conditions; they ensure that the string end-points do not travel in
any direction perpendicular to the D-branes, but they do not ensure
that the end-points are actually on the D-branes. Satisfying this
extra requirement has two consequences: firstly, it determines the
initial values ${\bf a}_{\perp}(0)$ and ${\bf b}_{\perp}(0)$, and
secondly --- and more importantly for our purposes --- it implies
that the curve ${\bf a}'(\xi)$ must satisfy the condition
\be\label{eq:avcond1}
 \frac{1}{2}\int_{-L}^{L} {\bf a}'_{\perp}(\xi){\rm d}\xi =
 {\bf \Delta}\ ,
\ee
where ${\bf \Delta}$ is a normal vector stretching from the lower brane at
$\sigma=L$ to the upper brane at $\sigma=0$. In other words, ${\bf \Delta}$
points in the $+z$-direction and has magnitude equal to the
inter-brane separation.

Finally, we have to fix the remaining symmetry, namely the space-time
symmetry corresponding to a free choice of space-time coordinates
$x^{\mu}$. In other words, we need to choose a unique inertial
frame. One obvious choice would be the zero momentum frame. This
choice is indeed possible, as one can check that the total momentum of
the string 
\be 
P^{\mu} \sim \int_{0}^{L} \dot{x}^{\mu}(t,\sigma){\rm  d}\sigma 
\ee 
is a time-like vector, and therefore there exists a Lorentz
transformation that would set the spatial part of this vector equal to
zero.  However, we need to avoid Lorentz transformations that involve
the Dirichlet directions, as these would change the configuration of
the D-branes. Therefore, the correct solution is to choose the frame
in which only the Neumann component of the momentum, ${\bf
  P}_{\parallel}$, is zero. In other words, we choose the unique frame
in which the momentum in directions parallel to the branes vanishes. 
Obviously, such a choice of frame will not affect the
configuration of the D-branes. It follows that in this frame, 
\be
\int_{0}^{L} \dot{\bf x}_{\parallel}(t,\sigma){\rm d}\sigma=0\ .
\label{cond}
\ee
Combining Eqs.~(\ref{eq:avcond1}) and (\ref{cond}) we get the condition
\be\label{eq:avcond2}
 \langle{\bf a}'\rangle = \frac{\bf \Delta}{L}\ ,
\ee
with the definition 
\be
 \langle{\bf v}\rangle\equiv \frac{1}{2L}\int_{-L}^{L} {\bf v}(\xi){\rm d}\xi~,
\ee
for any closed curve ${\bf v}(\xi)$ on the unit sphere, having
parameter length equal to $2L$. (This quantity can be geometrically
visualised as the average or centre-of-mass of such a curve.) Notice
that this condition has a corollary that we should have expected on
physical grounds: there is a lower bound on our choice of string
parameter length, $L\geq|{\bf \Delta}|$.

We have thus arrived at a geometric method for classifying solutions
for a system of a DBI string constrained by two parallel and
stationary D-branes. Working with a conformal and temporal gauge
parametrisation of the world-sheet, and also working in unique zero
Neumann-momentum space-time coordinates, we have found that any string
solution has a unique description in terms of one constant scalar,
$L$, two constant vectors, ${\bf a}_{\parallel}(0)$ and ${\bf
  b}_{\parallel}(0)$, as well as a single closed, parametrised curve
on the unit sphere, ${\bf a}'(\xi)$ (or, equivalently, ${\bf
  b}'(\xi)$), which satisfies the averaging condition
Eq.~(\ref{eq:avcond2}) (or an equivalent averaging condition for ${\bf
  b}'$).

\subsection*{Cusps}

To investigate cusps formation in DBI strings ending on two stationary
and parallel D$n$-branes, we will follow the same approach as for
Nambu-Goto string loops, we are more familiar with.  Actually, the
geometric visualisation of the DBI string solutions described here is
similar to that of closed Nambu-Goto string
loops~\cite{Vilenkin_shellard}, but with two important
differences. Firstly, whilst with closed string loops we are free to
choose both curves ${\bf a}'$ and ${\bf b}'$ independently, for a DBI
string ending on two stationary and parallel D$n$-branes we can only
choose one of these curves freely; the other one is automatically
determined by the inversion described by
Eqs.~(\ref{eq:Dbrane_boun1a}), (\ref{eq:Dbrane_boun1b}). Secondly,
with closed string loops we can always choose the zero momentum frame,
so that the average of the curves ${\bf a}'$ and ${\bf b}'$ can always
be constrained to the centre of the unit sphere. Here we can only
constrain the average of ${\bf a}'$ to a unique point within the unit
sphere determined by the separation and orientation of the D-branes,
${\bf \Delta}$. When we investigate the formation of cusps in this
system, we will see the consequences of the above mentioned
differences.

Cusps are found wherever the instantaneous velocity of a part of the
string reaches the speed of light. As with closed strings, it is
straightforward to show that cusps are present in DBI strings ending
in two stationary and parallel branes if, and only if, the curves
${\bf a}'$ and ${\bf b}'$ intersect on the unit sphere.  Therefore,
the prevalence of cusps in a string configuration is directly related
to the prevalence of intersections amongst pairs of closed curves on
the unit sphere, that satisfy a number of properties listed above.

For smooth closed string loops, where ${\bf a}'$ and ${\bf b}'$ are
independent smooth curves whose centres-of-mass are at the centre of
the unit sphere, it is very difficult for ${\bf a}'$ and ${\bf b}'$
curves to avoid intersection. Therefore, as it is usually said, closed
string loops have \emph{generically} cusps~\cite{turok1984}.
\begin{figure}
\begin{center}
 \includegraphics[scale=1.0]{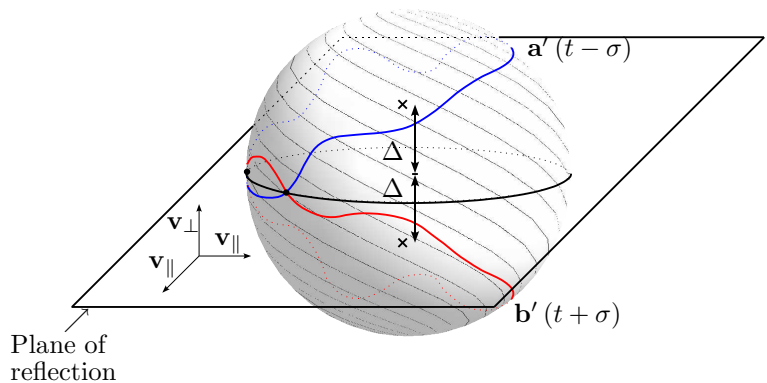}
 \caption{\label{fig1} The vectors ${\bf a}'\left(t-\sigma\right)$ and
 ${\bf b}'\left(t+\sigma\right)$  trace out closed curves on unit
 spheres that are separated by the inter-brane distance.}
\end{center}
\end{figure}
For the DBI string under consideration here, the curves are no
longer independent. The exact nature of their relationship to each
other depends on the dimensionality of the D$n$-branes. The important
case for cosmological implications is $n=1$ (D$1$-branes), which we discuss
in detail in the following sections. For $n=2$,
the two curves are reflections of each other across a plane passing
through the centre of the unit sphere (see, Fig.\ref{fig1}); for $n=1$ the two
curves are inversions of one another through a line passing through
the unit sphere; and for $n=0$ the
two curves are inversions of one another through the centre of the
unit sphere. Furthermore, the centres-of-mass of the two curves are
no longer necessarily located at the centre of the unit sphere -- their
locations are now governed by the inter-brane separation.

Even though we have managed to reduce the problem of cusps to a rather
simple-looking game involving intersections of curves satisfying
certain properties, it turns out to be very difficult to make any
quantitative predictions about this geometric puzzle. Nevertheless,
there are a few observations that are worth pointing out.

Firstly, for any $n$, the likelihood of cusps falls to zero as $L$
approaches the inter-brane separation distance $|{\bf \Delta}|$ (i.e.,
the centres-of-mass of the curves approach the surface of the unit
sphere). In this limit, the two curves, related by inversion, are
confined to shrinking antipodal regions of the sphere and the
likelihood of intersection becomes vanishingly small.

Secondly, for coincident D2-branes (with the DBI string having its
end-points on these coincident branes) there are always cusps. This is
because the two curves have centres-of-mass at the centre of the unit
sphere, and are related by reflections through a plane passing through
the centre of the unit sphere. The first condition forces each curve
to cross the plane of reflection at least twice, whilst the second
condition implies that the two curves intersect whenever they cross
the reflection plane. Assuming continuity, it is therefore reasonable
to expect that even for the case of non-coincident D2-branes, the
likelihood of cusps will grow in the limit $L\gg|{\bf \Delta}|$.

Thirdly, the situation is different for coincident D1- and D0-branes,
as although the averaging condition still forces the two curves to
pass through the same plane as before, whereas the inversion property
no longer implies that the two curves have to cross this plane at the
same point. Therefore, intersections are not guaranteed in the same
way as they are for coincident D2-branes.

\begin{figure}
\begin{center}
 \includegraphics[scale=0.75]{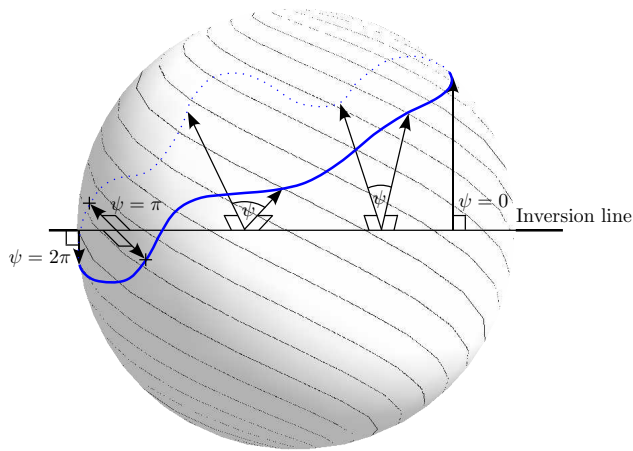}
 \caption{\label{fig2} When the line through which the two closed curves
are inverted pierces the closed curve, then there will generically be
and intersection of the two curves. This can be seen by considering the
two vectors perpendicular to the inversion line, ending on the curve.
When the angle between these vectors, $\psi$, is $\pi$ the inverted curve will
intersect the original curve (marked with crosses).}
\end{center}
\end{figure}

For D1-branes however we can see that the two curves will generically
intersect whenever the line though which they are inverted is enclosed
by the closed curves. To see this consider the two vectors
perpendicular to the inversion line that end on the closed
curve. Typically these will not be anti-parallel, however if they are,
then these points will be unaffected by the inversion and the ${\bf
  a}'$ and ${\bf b}'$ curves will intersect. If the inversion line is
enclosed by the closed curve, then at one extreme the two vectors will
have zero angle between them and at the other extreme this angle will
tend to $2\pi$. By continuity there must be a point at which the angle
between the two vectors is $\pi$ and hence we will get the
intersection between the ${\bf a}'$ and ${\bf b}'$ curves, necessary
for cusps formation (see, Fig.~\ref{fig2}).

\begin{figure}
\begin{center}
 \includegraphics[scale=0.75]{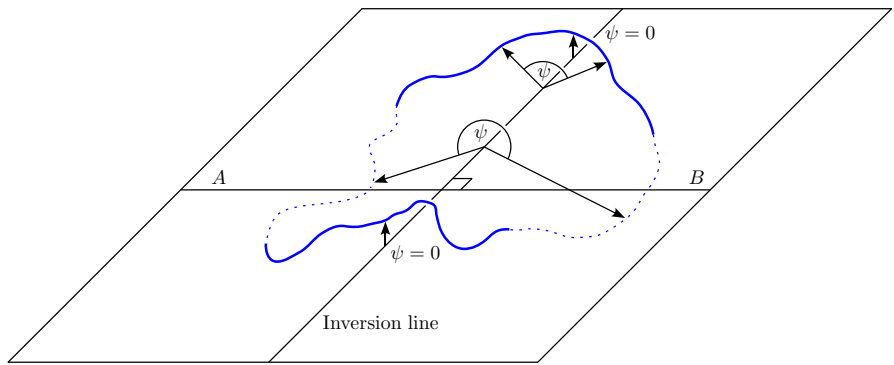}
 \caption{\label{fig3} If there exists a line that intersects the inversion
line at right angles, $AB$, that is topologically on the opposite side of
the closed curve, then there will be a pair of points that are invariant
under the inversion, i.e. there will be cusps. The sphere has not been
drawn for clarity.}
\end{center}
\end{figure}

\begin{figure}
\begin{center}
 \includegraphics[scale=1.0]{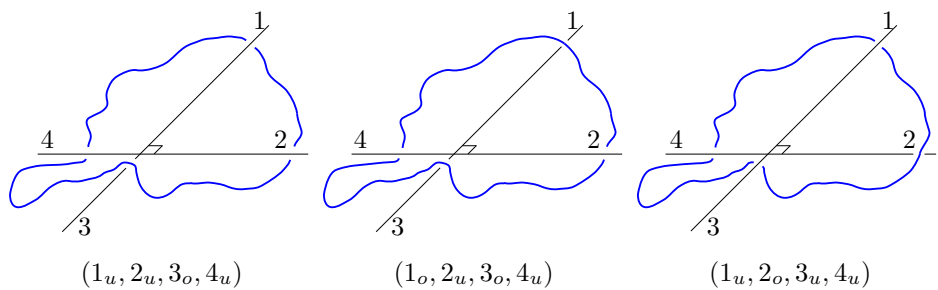}
 \caption{\label{fig:over_under} The legs of the inversion line are
labelled $(1,3)$ and the legs of the line intersecting it at right
angles are labelled $(2,4)$. We can then classify possible curves
by whether they go over or under each leg. Three examples are give above,
the first two will generically contain cusps, whilst the last will not.}
\end{center}
\end{figure}

We can extend the above, by considering closed curves that do not
encircle the inversion line, but for which there exists a line, $AB$,
that intersects the inversion line at ninety degrees and is
(topologically) on the opposite side of the closed curve (see,
Fig.~\ref{fig3}). To see that this scenario also has points on the
closed curve that are invariant under the inversion, consider the
plane containing the inversion line and the line $AB$. The closed
curve will have to intersect this plane at least four times (with the
possibility that two or more of these intersection points are
coincident in the extreme cases). Then as before we can construct the
vectors perpendicular to the inversion line that end on the closed
curve. The angle between these two vectors will be zero at both
extremes and because the line $AB$ in on the opposite side of the
closed curve to the inversion line, there will be at least one point
at which the angle between the two vectors is greater than $\pi$. By
continuity, there exists a pair of points on the closed curve at
which the angle between the vectors is $\pi$ and hence a pair of
points on the closed curve that are invariant under the inversion,
i.e. there will be cusps.
We can then classify all possible closed curves by considering a
general line intersecting the inversion line at right angles and
describing the closed curve by whether it goes over or under each leg
of the resulting cross (see, Fig.~\ref{fig:over_under}). If we label the
legs of the cross on the inversion line by $1,3$ and the remaining
legs of the cross by $2,4$ then the following curves have cusps,
\be\label{eq:cusps_num}
\begin{array}{ccc}
 \left( 1_{\bf o}, 2_{\bf u}, 3_{\bf u}, 4_{\bf u} \right), \ \ & 
\left(1_{\bf o}, 2_{\bf o}, 3_{\bf u}, 4_{\bf u}\right),
\ \ & \left(1_{\bf o}, 2_{\bf u}, 3_{\bf o}, 4_{\bf u}\right),  \\
 \left(1_{\bf u}, 2_{\bf u}, 3_{\bf o}, 4_{\bf u}\right) ,\ \ & 
\left(1_{\bf o}, 2_{\bf u}, 3_{\bf u}, 4_{\bf o}\right),\ \ & \\
\end{array}
\ee as well as their reflections ${\bf o}\leftrightarrow {\bf u}$,
where the subscripts indicate whether the curve went over $X_{\bf o}$,
or under $X_{\bf u}$ the leg $X$.
The first and second groups in Eq.~(\ref{eq:cusps_num}) have cusps
because the inversion line is encircled by the closed curve, whilst
the right-hand most curve has cusps by the generalisation given
above. In total there are another three curves (and their reflections)
that will not produce cusps, $\left(1_{\bf o}, 2_{\bf o}, 3_{\bf o},
4_{\bf o}\right)$, $ \left(1_{\bf u}, 2_{\bf o}, 3_{\bf u}, 4_{\bf
  u}\right)$ and $\left(1_{\bf u}, 2_{\bf u}, 3_{\bf u}, 4_{\bf
  o}\right)$.  If each of these possibilities are equally likely,
which is a reasonable assumption if the centre of the mass of the
closed curve and the centre of the cross are at the origin, then we
would expect to have cusps in more than half of curves. As the centre
of mass of the closed curve is moved away from the origin, the
probability of each of these curves would no longer be equal. In
particular as the closed curves become restricted to shrinking
antipodal region we can see that the $\left(1_{\bf o}, 2_{\bf o},
3_{\bf o}, 4_{\bf o}\right)$ curve (and its reflection) would become
increasingly more likely, thus reducing the probability of having
cusps, in line with our earlier expectations.

Thus, we have shown that whilst cusps may not be a generic feature of a
string stretched between two D1-branes, when $|\Delta |\ll L$ we would
expect to find cusps in a significant fraction of cases. In the
following we will need a slight generalisation of the above
proofs. 

\begin{figure}
\begin{center}
 \includegraphics[scale=1.0]{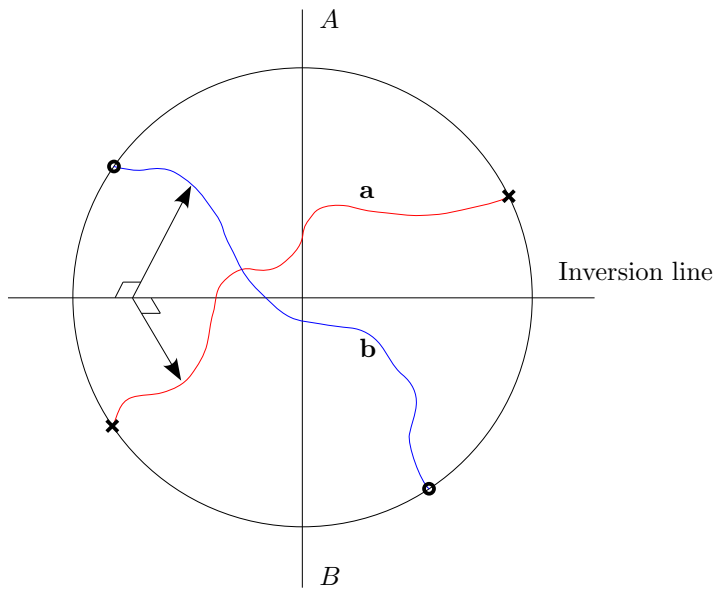}
 \caption{\label{fig:deform} If a closed curve ${\bf a}$ intersects the plane
made up of the inversion line and a line perpendicular to it as shown then it
will encircle both lines. The inversion of ${\bf a}$ through the inversion line
results in a curve that intersects the plane through two opposite 
quadrants. For any deformation of this inverted closed curve, ${\bf b}$,
that respects the positions of these intersection, there will be a point
at which the two curves ${\bf a}$ and ${\bf b}$ intersect. In the diagram a
top down view of the plane and unit sphere is given, with the curves ${\bf a}$
and ${\bf b}$ extending out of the page.}
\end{center}
\end{figure}

For the cases of the closed curve encircling the inversion line, we
have seen that the closed curve will intersect its inversion, here we
extend this to deformations of this inverted curve. In particular,
consider a closed curve ${\bf a}$ that encircles both the inversion
line and a line $AB$ intersecting the inversion line at right angles
(the two curves in the second column of Eq.~(\ref{eq:cusps_num})).
The inversion of ${\bf a}$ will yield a closed curve ${\bf b}$ that
also encircles both these lines, but with the opposite
orientation. Any subsequent deformation of ${\bf b}$ (or indeed ${\bf
  a}$) that preserves the fact that the curve encircles both lines,
will result in intersections between the two closed curves. To see
this consider the plane consisting of the inversion line and $AB$. We
arbitrarily define one side of this plane to be positive, then we have
for the closed curves ${\bf a}$ and ${\bf b}$ to correctly encircle
the inversion line and the line $AB$, that they must intersect the
plane twice each, in opposite quadrants as in Fig.~\ref{fig:deform}.
The angle between the two vectors perpendicular to the inversion line
(or $AB$) and ending on ${\bf a}$ and ${\bf b}$ must at some point
reach zero, which implies the closed curves intersect. This holds
equally for the negative side of the plane. Essentially what this says
is that the result for closed curves and their inversions through a
line is a topological one and holds equally well for subsequent
deformations of either closed curve that preserves the topological
relations between the closed curves and the inversion line. The
requirement that the relation between the line perpendicular to the
inversion line and the closed curves is there to ensure that the
subsequent deformation does not `undo' the inversion.

%%%%%%%%%%%%%%%%%%%%%%%%%%%%%%%%%%%%%%%%%%%%%%%%%%%%%%%%%%%%%%%%%%%%%%%%%%%

\section{Three string junctions}\label{sec:3string}

Let us consider the simplest DBI string system that contains a
junction: that of three DBI strings joined at a single junction.  For
the sake of simplicity, we attach the free ends of the strings (the
ends that are not connected to the junction) to three flat, stationary
and parallel D$n$-branes, which (without loss of generality) are
spatially separated in the $z$-direction. In what follows, only the
boundary conditions for the three DBI strings associated with these
D$n$-branes are important and we need not be concerned with the
precise nature (dimensionality) of the branes themselves\footnote{This
  will be important later on, when we discuss (as an example) the case
  of a three string junction composed by an F-,D-string, and their
  FD-string bound state.}.

We label the three strings with an index $i=1,2,3$. As in
Section~\ref{sec:Dbranes}, we parametrise each string's world-sheet,
$x_i^{\mu}(\tau,\sigma)$, so that $\sigma=0$ corresponds to the end of
the string attached to a D$n$-brane, whilst $\sigma=L_{i}(\tau)$
corresponds to the three-string junction. Thus, $L_{i}(\tau)$ is the
parameter length of the $i^{\mathrm{th}}$-string at world-sheet time
$\tau$.  We denote the world-line of the junction by
$\bar{x}^{\mu}(\tau)$ and the (single-component) gauge field present
on the junction by $\bar{A}(\tau)$.

We exploit the re-parametrisation invariance of each string
world-sheet to work in conformal gauge. In this gauge, the action for
this configuration of three DBI strings reads~\cite{Copeland:2007nv}
\begin{eqnarray}\label{eq:3stringaction} 
S &=& - \sum_i \mu_i \int{\rm d}\tau\int_{0}^{L_{i}(\tau)}{\rm
  d}\sigma \sqrt{ -{x'_i}^2 {\dot{x}_i}^2 -\lambda
  (F^i_{\tau\sigma})^2 } \nonumber \\ 
%&&\hspace{-3.5truecm}
&&+ \sum_i \int{\rm d}\tau
\left\{{\bf f}_i(\tau) \cdot [
  x_i(\tau,L_i(\tau))-\bar{x}(\tau)]\right. \nonumber \\ 
&&\ \ \ \ \ \ \left. +g_i(\tau)\left[A^i_{\tau}(\tau,L_i(\tau))+ \dot{L}_i A^i_\sigma
  (\tau,L_i(\tau)) - \bar{A}(\tau) \right]\right\}~,
\end{eqnarray}
where $\mu_i$ is the tension of the $i^{\mathrm{th}}$-string, which
for a ($p$,$q$)-string is given by $\mu_i=|q_i|/(g_s \lambda)$, and
$A^{i}_{\alpha}$ and $F^{i}_{\alpha\beta}$ are the gauge field and
gauge field strengths respectively on the
$i^{\mathrm{th}}$-string. The Lagrange multipliers ${\bf f}_i(\tau)$
constrain the strings to meet at the junction $\bar{x}(\tau)$, whilst
the Lagrange multipliers $g_i(\tau)$ impose to the component of each
string's gauge field that is tangential to the junction world-line to
coincide with the junction gauge field $\bar{A}(\tau)$.

We then derive the equations of motion for this action\footnote{This was
first done in Ref.~\cite{Copeland:2007nv}; however we shall
include the results here for the sake of clarity and completeness.}.
By varying the action with respect to ${x_i}^{\mu}(\tau,\sigma)$ and
$A^{i}_{\alpha}$, we find that the equations of motion for the string
world-sheets are:
\be
\ddot{x}_i^{\mu}-{x_i^{\mu}}''=0~,
\ee
and the equations of motion for the gauge fields once again imply the
conservation of electric flux along each string,
i.e., $\partial_{\tau}p_i=\partial_{\sigma}p_i=0$, where
\be
p_i=\frac{\lambda^{2}\mu_{i}F^i_{\tau\sigma}}
 {\sqrt{-{x'_i}^2 {\dot{x}_i}^2 -\lambda
    \left(F^i_{\tau\sigma}\right)^2}}~.
\ee
The above variations of the action also lead to certain boundary
conditions at the junction which, when combined with the equations
obtained by varying the action with respect to $\bar{x}(\tau)$ and
$\bar{A}(\tau)$, give the following conservation laws:
\be\label{eq:juncbc1}
\sum_i\bar{\mu}_i\left({x_i^{\mu}}'+\dot{L}_{i}\dot{x}_i^{\mu}\right)=0
\ee
and
\be\label{eq:fstringcons}
\sum_{i}p_{i}=0\ ,
\ee
where the effective string tension $\bar{\mu}_i$ is 
\be\label{eq:mubardef}
\bar{\mu}_i=\sqrt{\lambda^{2}\mu_{i}^{2}+p_i^2}
           =\sqrt{\frac{q_i^2}{g_s^2}+p_i^2}\ .
\ee
Finally, variation of the action with respect to the Lagrange multipliers,
$f_i(\tau)$ and $g_i(\tau)$, leads to the required constraints
\be\label{eq:juncbc2}
x_i\left(\tau,L_i(\tau)\right)=\bar{x}\left(\tau\right)
\ee
and
\be
A^i_{\tau}\left(\tau,L_i(\tau)\right)+ \dot{L}_i A^i_\sigma
   \left(\tau,L_i(\tau)\right) = \bar{A}(\tau)~.
\ee
The action given in Eq.~(\ref{eq:3stringaction}) can only represent a
network of ($p$,$q$)-strings when none of the $q_i$'s are zero. In the
presence of a ($p$,0)-string, this action has to
be modified by replacing the corresponding DBI kinetic term by a
Nambo-Goto (gauge-free) kinetic term with string tension
$\mu_i=p$. Nevertheless, it is straightforward to show that the
equations of motion, boundary conditions and conservation equations,
derived from this modified action, agree with those given above for
the original action.

Furthermore, when we are dealing with a system of ($p$,$q$)-strings
(rather than generic DBI strings, which can have arbitrary tensions),
we can show that the strings must satisfy the additional
conservation condition~\cite{Copeland:2007nv}
\be
\sum_{i}q_{i}=0\ ,
\ee
which prevents any of the D-strings contained in the three ($p$,$q$)
bound states from ending at the junction. [Equation
(\ref{eq:fstringcons}) has much the same consequence for the
F-strings.]

Finally, as mentioned in the previous section, the conformal gauge condition
admits a residual re-parametrisation invariance, which can be fixed by
providing a supplementary gauge condition. Once again, one can show
that temporal gauge,
\be
\tau=t,\quad x_i^{\mu}(t,\sigma)=\left(t,{\bf x}(t,\sigma)\right)
 \quad\mathrm{and}\quad\bar{x}^{\mu}(t)=\left(t,\bar{\bf x}(t)\right)~,
\ee
is both consistent with the equations of motion, and also completely fixes the
world-sheet parametrisation.

We note that the above equations of motion, the boundary conditions
and the conservation laws combine to give us a system that exhibits
self-duality under the S-duality transformation
\be
p\leftrightarrow q \quad\mathrm{and}\quad g_{s}\to\frac{1}{g_s}\ .
\ee
This shall become important later, when we come to discuss the
properties of this system in the weak coupling limit $g_s \ll 1$;
any conclusion we make about the behaviour of the system in this regime
also applies in the strong coupling limit $g_s \gg 1$, provided that
we interchange the roles of the F- and D-strings.

We shall now try to classify the string solutions of this system in a
geometric manner, by attempting to follow steps similar to those taken
in Section \ref{sec:Dbranes}.

In conformal and temporal gauge, we have the usual
wave equation for each string world-sheet
\be
 \ddot{\bf x}_i - {\bf x}_i'' = 0~.
\ee
This admits the general solution
\be\label{eq:3sgensol}
 {\bf x}_i = \frac{1}{2}\left[ {\bf a}_i\left(t-\sigma\right)
             + {\bf b}_i\left( t+\sigma \right) \right]~,
\ee
for arbitrary single-variable functions ${\bf a}_i$ and ${\bf b}_i$,
subject to the conformal gauge constraints
\be\label{eq:gaugecon3}
 |{\bf a}'_i|^2 = |{\bf b}'_i|^2 = 1~.
\ee
As with the previous single-string system, we must investigate how the
boundary conditions on the three strings affect our {\sl hitherto} free
choice in selecting the two curves on the unit sphere that ${\bf
  a}'_i$ and ${\bf b}'_i$ represent.

The D$n$-brane boundary conditions at the $\sigma=0$ end of each
string are the most straightforward; they provide the following
conditions: 
\beq\label{eq:3string_boun1a} 
{\bf a}'_{i\parallel}(t) &=& {\bf b}'_{i\parallel}(t)~, \\ {\bf
  a}'_{i\perp}(t) &=& - {\bf
  b}'_{i\perp}(t)~,\label{eq:3string_boun1b} 
\eeq 
which are analogous to Eqs.~(\ref{eq:Dbrane_boun1a}) and
(\ref{eq:Dbrane_boun1b}).

We next consider the implications of the boundary conditions at the
junction, given by Eqs.~(\ref{eq:juncbc1}) and (\ref{eq:juncbc2}),
which in temporal gauge are more conveniently expressed as:
\beq
\label{eq:juncbc1a}
 \sum_i \bar{\mu}_i \dot{L}_i &=&0~,\\
\label{eq:juncbc1b}
 \sum_i \bar{\mu}_i
    \left({\bf x}'_i + \dot{L}_{i}\dot{\bf x}_i\right)&=&0~,\\
\label{eq:juncbc2a}
 {\bf x}_i^{\mu}\left(t,L_i(t)\right)
  &=& {\bf x}_j^{\mu}\left(t,L_j(t)\right)\quad~,\ \ \ \ \forall~i,j~.
\eeq
By substituting the general solution, Eq.~(\ref{eq:3sgensol}), into
Eq.~(\ref{eq:juncbc1b}) and the $t$-derivative of
Eq.~(\ref{eq:juncbc2a}), we obtain three independent conditions, given
by
\begin{eqnarray}\label{eq:3string_boun2}&&
\hspace{-2.5truecm}
\left[\bar{\mu}_1+\bar{\mu}_2+\bar{\mu}_3\right]\left[
1+\dot{L}_1\right]{\bf b}'_1\left(t+L_1(t)\right) =\nonumber
\\
&& 
\hspace{-2.5truecm}\qquad\qquad\qquad
 \left[\bar{\mu}_1-\bar{\mu}_2-\bar{\mu}_3\right]
   \left[1-\dot{L}_1\right]{\bf a}_1'\left(t-L_1(t)\right)\nonumber
\\
&&
\hspace{-2.5truecm}
\qquad\qquad\qquad
 +2\bar{\mu}_2\left[1-\dot{L}_2\right]{\bf a}_2'\left(t-L_2(t)\right)
 +2\bar{\mu}_3\left[1-\dot{L}_3\right]{\bf a}_3'\left(t-L_3(t)\right)~,
\end{eqnarray}
and its two counterparts under cyclic permutation of the string label
indices.

Following the same methodology as in Section \ref{sec:Dbranes}, we
look for necessary conditions that need to be satisfied by curves that
are subject to the above constraints, as well as to those we obtain by
squaring the above equations and imposing the gauge constraints,
Eq.~(\ref{eq:gaugecon3}).  By doing so, we arrive at a system of three
simultaneous equations which are polynomial in the three
$\dot{L}_i$. We can ``solve'' this system of simultaneous equations,
in the sense that we can re-arrange them to give a set of expressions
for $\dot{L}_i$. However, since the coefficients in these simultaneous
polynomial equations involve scalar products between the various ${\bf
  a}'_i(t-L_{i}(t))$, which themselves depend on $\dot{L}_i$, these
expressions for $\dot{L}_i$ will in general be ordinary differential
equations in $L_i(t)$.  Therefore, by inverting this system of
equations, we find seven independent solutions; six of those are
\be\label{eq:solspec}
\dot{L}_1=\pm 1~,\quad\dot{L}_2=\pm 1~,\quad\dot{L}_3=
  \mp\frac{\bar{\mu}_1+\bar{\mu}_2}{\bar{\mu}_3}~,
\ee
and cyclic permutations. Notice that these solutions happen to be
independent of ${L}_i$, and therefore can be integrated
directly. The seventh solution however does depend on ${L}_i$, via
the scalar products
\be
c_{ij}={\bf a}'_i(t-L_i(t))\cdot{\bf a}'_j(t-L_j(t))~,
\ee
and is given by
\be
\label{eq:sols}
\frac{\bar{\mu}_1}{\bar{\mu}_1+\bar{\mu}_2+\bar{\mu}_3} \left(
1-\dot{L}_i\right)= \frac{ M_1 \left( 1-c_{23}\right)}{M_1\left(
  1-c_{23}\right)+ M_2\left( 1-c_{13}\right)+ M_3\left(
  1-c_{12}\right)}~,
\ee
with cyclic permutations giving expressions for $\dot{L}_2$ and
$\dot{L}_3$, where
\be
M_1 = \bar{\mu}_1^2 - \left(\bar{\mu}_2 - \bar{\mu}_3 \right)^2~,
\ee
with cyclic permutations giving $M_2$ and $M_3$, respectively.

Let us briefly examine the first six solutions. By substituting these
solutions back into Eq.~(\ref{eq:3string_boun2}), we find that three
of them require that
\be
 {\bf a}'_{1}(t-L_1)={\bf a}'_{2}(t-L_2)={\bf a}'_{3}(t-L_3)~,
\ee
whilst the other three require that
\be
 {\bf b}'_{1}(t+L_1)={\bf b}'_{2}(t+L_2)={\bf b}'_{3}(t+L_3)~.
\ee
The seventh solution, Eq.~(\ref{eq:sols}), gives us greater freedom in
choosing ${\bf a}'_i(t-L_i)$ and ${\bf b}'_i(t+L_i)$ -- that is, for
\emph{any} given set of ${\bf a}'_i(t-L_i)$ (as long as they are not
all equal), we can calculate $\dot{L}_i$ using Eq.~(\ref{eq:sols}) and
${\bf b}'_i(t+L_i)$ using Eq.~(\ref{eq:3string_boun2}).

Furthermore, consider calculating $\dot{L}_i$ using
Eq.~(\ref{eq:sols}) for initially unequal ${\bf a}'_i(t-L_i)$, and
subsequently taking the limit where ${\bf a}'_i(t-L_i)$ become
equal. The values of $\dot{L}_i$ in this limit are, in general,
different from the values of $\dot{L}_i$ suggested by
Eq.~(\ref{eq:solspec}) for exactly equal ${\bf a}'_i(t-L_i)$. In other
words, there is no way of continuously deforming a string solution
that is described by Eq.~(\ref{eq:sols}) into one of the special
solutions described by Eq.~(\ref{eq:solspec})\footnote {We note that
  in Ref.~\cite{Copeland:2006eh} only Eq.~(\ref{eq:sols}), and its
  cyclic permutations, have been discussed as string solutions.}.

Combining the two results above, we conclude that the special string
solutions characterised by Eq.~(\ref{eq:solspec}) form a disconnected
and effectively lower dimensional part of the space of
solutions. Thus, when we discuss \emph{generic} properties of this
system, we may safely neglect these special solutions as, although
they are perfectly valid as dynamical solutions, they take up an
effectively zero-volume portion of the total space of solutions.

Finally, notice that the conditions, Eq.~(\ref{eq:3string_boun1a}),
(\ref{eq:3string_boun1b}), (\ref{eq:juncbc1a}) and
Eq.(\ref{eq:3string_boun2}) and its two counterparts under cyclic
permutation, are not quite enough to fully enforce the Dirichlet and
junction boundary conditions.  Equation (\ref{eq:3string_boun1b})
forces the non-junction ends of the strings to move tangentially to
the D$n$-branes, but in order to ensure that these ends actually meet
the branes, we need to impose the extra conditions
\be 
{\bf x}_{i\perp}(0,0)={\bf D}_i~, 
\ee 
where ${\bf D}_i$ is the position vector of the point where the
D$n$-brane, attached to the $i^{\mathrm{th}}$-string, intersects the
$z$-axis. Similarly, Eq.~(\ref{eq:3string_boun2}) and its cyclic
permutations force the junction ends of the string to move in tandem
with one another, but in order to ensure that these ends actually meet
at the junction, we need to impose the extra conditions 
\be 
{\bf x}_{i}(0,L_i(0))={\bf x}_{j}(0,L_j(0))\quad\ \ ,\ \ \ \forall~i,j~.  
\ee 
It is straightforward to check that the effect of enforcing these
conditions (apart from fixing the constants ${\bf a}_{i\perp}(0)$ and
${\bf b}_{i\perp}(0)$, which are inconsequential for our purposes) is
to introduce the following averaging constraints on the curves ${\bf
  a}_i$: 
\be
\label{eq:3Davcond} 
\int_{-L_i(0)}^{L_i(0)} {\bf a}'_{i\perp}(\xi) {\rm d}\xi -
\int_{-L_j(0)}^{L_j(0)} {\bf a}'_{j\perp}(\xi) {\rm d}\xi = {\bf
  \Delta}_{ij}\quad\ \ ,\ \ \ \forall~ i,j~, 
\ee 
where 
\be 
{\bf \Delta}_{ij} = {\bf D}_i - {\bf D}_j ~,
\ee 
is the normal separation vector from the $i^{\mathrm{th}}$ to the
$j^{\mathrm{th}}$ D$n$-brane.

Finally, as discussed in Section \ref{sec:Dbranes}, we fix the
remaining space-time symmetry in this system by choosing the unique
inertial frame in which the Neumann component total momentum of the
system, $P_{\parallel}$, is zero. This leads to the averaging
condition
\be\label{eq:3Pavcond}
\sum_i \int_{-L_i(0)}^{L_i(0)} {\bf a}'_{i\parallel}(\xi) {\rm d}\xi
 = 0~.
\ee

At the end of this long process, we have succeeded in classifying
dynamical solutions of the DBI string system containing a junction in
terms of the curves on the unit sphere ${\bf a}'_i(t-\sigma)$ and
${\bf b}'_i(t+\sigma)$, the scalar functions $L_i(t)$ and the
constants ${\bf a}_i(0)$ and ${\bf b}_i(0)$. In particular, we have
found that any choice of these quantities, subject to the constraints
given by Eqs.~(\ref{eq:3string_boun1a}), (\ref{eq:3string_boun1b}),
(\ref{eq:3string_boun2}), (\ref{eq:sols}), (\ref{eq:3Davcond}) and
(\ref{eq:3Pavcond}), corresponds to a unique string solution (and vice
versa).

Although most of these constraints have clear geometrical
interpretations (as it was the case in Section \ref{sec:Dbranes}), the
constraints represented by Eqs.~(\ref{eq:3string_boun2}) and
(\ref{eq:sols}) are not at all easy to visualise for general string
tensions $\bar{\mu}_i$.  Moreover, as cusps on these strings are
associated with intersections of the curves ${\bf a}'_i$ and ${\bf
  b}'_i$, we need a full geometric picture of these constraints in
order to consider the likelihood of cusps in this system. Fortunately
however, we shall find that progress can be made in the case of
($p$,$q$)-string junctions for certain limits of the coupling
constants.

%%%%%%%%%%%%%%%%%%%%%%%%%%%%%%%%%%%%%%%%%%%%%%%%%%%%%%%%%%%%%%%%%%%%%%%%%%%
\subsection*{Example: An F-, D-, FD-string junction}

Suppose that the string labelled by ``1'' is a (1,0)-string
(F-string); the string labelled by ``2'' is a (0,1)-string (D-string)
and the string labelled by ``3'' is a (1,1)-string (FD-string).  Using
Eq.~(\ref{eq:mubardef}) we can expand the effective tensions of these
strings as a series in the perturbative string coupling, $g_{\rm s}$:
\be
\bar{\mu}_1 = 1~, \ \ \ \bar{\mu}_2 = \frac{1}{g_{\rm s}}~,
\ \ \ \bar{\mu}_3 = \sqrt{ 1+\frac{1}{g_{\rm s}^2} } = \frac{1}{g_{\rm s}} +
\frac{g_{\rm s}}{2} + {\cal O}\left( g_{\rm s}^3\right)~. 
\ee 
Therefore, for $g_{\rm s} \ll 1$, we can write down a perturbative
expansion for ${\bf a}'_i$: \be {\bf a}'_i={{\bf a}'_i}^{\sss
  (0)}(\xi)+g_s {{\bf a}'_i}^{\sss (1)}(\xi)+ g_s^2{{\bf a}'_i}^{\sss
  (2)}(\xi)+\cdots 
\ee
and similarly for ${\bf b}'_i$ and $\dot{L}_i$, and substitute these
expansions into the boundary conditions given in
Eqs.~(\ref{eq:3string_boun2}) and (\ref{eq:sols}), again fixing the 
boundary conditions at the $\sigma_i = 0$ ends of the three strings 
by attaching them to parallel D branes. By doing so, and
matching coefficients order-by-order, we find the following boundary
conditions for the leading order terms: 
\beq \label{eq:b1pert}
\left(S_{23}^{\sss (0)}- 2S_{13}^{\sss (0)}-2S_{12}^{\sss (0)}\right)
     {{\bf b}'_1}^{\sss (0)} &=& S_{23}^{\sss (0)} {{\bf a}'_1}^{\sss
       (0)} -2S_{13}^{\sss (0)}{{\bf a}'_2}^{\sss (0)} -2S_{12}^{\sss
       (0)}{{\bf a}'_3}^{\sss (0)} ~,\\
\label{eq:b2a3}
{{\bf b}'_2}^{\sss (0)} &=& {{\bf a}'_3}^{\sss (0)} ~,\\
\label{eq:b3a2}
{{\bf b}'_3}^{\sss (0)} &=& {{\bf a}'_2}^{\sss (0)} ~,
\eeq
and
\be\label{eq:L1dotpert}
\dot{L}_1^{\sss (0)}=1-
 \frac{S_{23}^{\sss (0)}}{S_{12}^{\sss (0)}+S_{13}^{\sss (0)}}
  ~,\quad
\dot{L}_2^{\sss (0)}=
 \frac{S_{12}^{\sss (0)}-S_{13}^{\sss (0)}}{S_{12}^{\sss (0)}+S_{13}^{\sss (0)}}
  ~,\quad
\dot{L}_3^{\sss (0)}=
 \frac{S_{13}^{\sss (0)}-S_{12}^{\sss (0)}}{S_{12}^{\sss (0)}+S_{13}^{\sss (0)}}
  ~,
\ee
where $S_{ij}=\frac{1}{2}(1-c_{ij})$ and, for the sake of clarity, we
have used the short-hand
\be 
{\bf a}'_i = {\bf a}'_i(t-L_i(t))\quad\mathrm{and}\quad
{\bf b}'_i = {\bf b}'_i(t+L_i(t))~.
\ee
Being interested only to the leading-order behaviour of the strings,
we shall henceforth drop the superscripts ${}^{\sss (0)}$ in the
discussion on the perturbative dynamics of the system.

The above equations have a highly intuitive physical
interpretation, which is best seen by rewriting Eqs.\ (\ref{eq:b2a3})
and (\ref{eq:b3a2}) in the form
\be
\dot{\bf x}_2(t,L_2(t))=\dot{\bf x}_3(t,L_3(t))\quad\mathrm{and}\quad
{\bf x}'_2(t,L_2(t))=-{\bf x}'_3(t,L_3(t))~.
\ee
In other words, as $g_{\rm s}\to 0$, the D-string and FD-string
effectively become one {\sl continuous} string of constant (overall)
length, $L_2(t)+L_3(t)$, obeying the usual equation of motion, which
is unaffected by the dynamics of the much lighter F-string.

The dynamics of the F-string are then determined by the boundary
conditions, Eqs.\ (\ref{eq:b1pert}) and (\ref{eq:L1dotpert}), imposed
by the combined string (composed by the D- and the FD-string) on the
$\sigma=L_1(t)$ end of the F-string. Therefore, as far as the F-string
is concerned, the combined string is effectively a D1-brane (with its
own prescribed motion) on which the F-string produces no
back-reaction. These boundary conditions may be better understood when
expressed in terms of the orientation, ${\bf x}'_2(t,L_2(t))$, and
velocity, $\dot{\bf x}_2(t,L_2(t))$, of the effective D1-brane, as:
\be\label{eq:b1pert2}
\left[|{\bf x}'_2|^{2}-2(1-{\bf a}'_{1}\cdot\dot{\bf x}_2)\right] {\bf
  b}'_1 = -|{\bf x}'_2|^{2}\mathcal{R}{\bf a}'_1- 2(1-{\bf
  a}'_{1}\cdot\dot{\bf x}_2)\dot{\bf x}_2~, \ee and
\be\label{eq:L1dotpert2} \dot{L}_1= \frac{|\dot{\bf x}_2|^{2}-{\bf
    a}'_{1}\cdot\dot{\bf x}_2}{1-{\bf a}'_{1}\cdot\dot{\bf x}_2}~, \ee
where \be \mathcal{R}{\bf a}'_{1} = -{\bf a}'_{1}+ \frac{2({\bf
    a}'_{1}\cdot{\bf x}'_2){\bf x}'_2}{|{\bf x}'_2|^2}~, 
\ee 
is a linear transformation that inverts ${\bf a}'_1$ through the line
$\{\lambda{\bf x}';\lambda\in\mathbb{R}\}$, which is parallel to the
combined string and passes through the origin of the unit sphere.

After a little thought, it is apparent that Eq.~(\ref{eq:b1pert2})
causes ${\bf b}'_1$ to lie on the semi-circle that is obtained by
projecting the line $\{\mathcal{R}{\bf a}'_1 + \lambda\dot{\bf
  x}_2;\lambda\in\mathbb{R}\}$ onto the unit sphere ($\dot{\bf x}_2$
is perpendicular to ${\bf x}'_2$ by the conformal gauge
conditions). However, the exact position of ${\bf b}'_1$ on this
semi-circle depends on both the magnitude, $|\dot{\bf x}_2|$, and the
angle that $\dot{\bf x}_2$ makes with ${\bf a}'_1$, in a rather
complicated manner. Nevertheless, it is possible to make further
progress in two interesting limits: firstly, when the combined string
is moving slowly at the junction ($|\dot{\bf x}_2| \ll 1$); and
secondly, when the combined string is moving highly relativistically at
the junction, ($|{\bf x}'_2| \ll 1$). In both limits, we follow a
perturbative procedure as before by expanding Eqs.~(\ref{eq:b1pert2})
and (\ref{eq:L1dotpert2}) in the relevant small parameter and
extracting the leading order behaviour.

When the combined string is moving slowly ($|\dot{\bf x}_2| \ll 1$),
we obtain the following leading order boundary conditions:
\be \label{eq:condition1}
{\bf b}'_1 = \mathcal{R}{\bf a}'_1 \quad\mathrm{and}\quad \dot{L}_1 =
0~, 
\ee 
which are exactly the Dirichlet boundary conditions that we derived in
Section \ref{sec:Dbranes} for a single DBI string attached to a
non-dynamical D1-brane at its $\sigma=L$ end. 

As we saw in Section~\ref{sec:Dbranes}, a string ending on D$1$-branes
will have cusps a significant fraction of the time. Here, we have found
that precisely the boundary conditions necessary for these cusps to
form, occur for an F-string at a three string junction, in the limit
$|\dot{\bf x}_2| \ll 1$. We have taken the $\sigma=0$ end of the
F-string to have boundary conditions associated with a D-brane, which
can now be replaced with a second three string junction. Thus, we see
that cusps would be expected to form on F-strings ending on two three
string junctions, in the limit that the D/FD-string is moving
slowly. This can be extended by noting that for a general $\dot{\bf
  x}_2$ we have a deformation of a pure inversion. In particular, a
point on the curve ${\bf a}_1'$ is inverted through ${\bf x}_2'$ and
then {\it pulled} along the semi-circle made up of the inverted point
and $\pm \widehat{\dot{\bf x}_2}\equiv \dot{\bf x}_2/ |\dot{\bf
  x}_|$. In Section~\ref{sec:Dbranes}, we showed that, if we consider
the plane containing the lines ${\bf x}_2'$ and $\dot{\bf x}_2$, then
a curve ${\bf a}_1'$ that encircles both lines, i.e. pierces the plane
in opposite quadrants, will intersect the curve ${\bf b}_1'$ provided
that ${\bf b}_1'$ also pierces the plane in opposite quadrants. By
considering the projection of Eq.~(\ref{eq:b1pert2}) onto both ${\bf
  x}_2'$ and $\dot{\bf x}_2$, it is easily shown that
\be 
{\rm sign}\left( {\bf b}_1'\cdot {\bf x}_2' \right) = {\rm sign}\left(
{\bf b}_1'\cdot \dot{\bf x}_2 \right)~.  
\ee 
Thus, we see that indeed, ${\bf b}_1'$ does intersect the plane in
opposite quadrants and hence will intersect ${\bf a}_1'$ in a
significant proportion of cases.

It is also worth mentioning that the $|\dot{\bf x}_2|\ll 1$, or static
brane limit, is the approximation taken in the quatisation of standard
string theory. Since we are working with the DBI action,
Eq.~(\ref{eq:3stringaction}), which is derived as a low energy limit
of string theory, we should for consistency restrict ourselves to this
situation. The action given by Eq.~(\ref{eq:3stringaction}) can be
taken as a prototype for string junctions away from this limit,
however in this case the motivation from string theory becomes less
clear.

In addition, we wish to consider the phenomenology of, for example,
brane inflation, in which it is expected that networks of F- and D-
strings are produced, which can go on to form junctions. These two
networks would be produced at the same energy scale and because the
are independent we would generically expect the heavy D-strings to be
moving slower than the light F-strings. Thus, the limit $|\dot{\bf
  x}_2|\ll 1$ would be satisfied early in the evolution of the
networks. As the dynamics of the strings become coupled via the
formation of junctions, it is no longer clear that this would be the
case and simulations would be required to estimate when this
approximation breaks down. However, as we have shown, cusps will remain
significant away from this simplifying limit.

We find that a junction formed from an F-, D- and FD-string behaves as
an F-string ending on a static (locally) straight D-string. We have
restricted our attention to the case where the $\sigma_i = 0$ ends of
the strings are attached to a D-brane, however if we take this D-brane
to be a D-string, then our results show that it could be replaced with
another F-string, D-string and FD-string junction. Thus, an F-string
stretched between two three-string junctions behaves as an F-string
between two D1-branes, to order $g_s$. In particular,
Eq.~(\ref{eq:condition1}) implies that the coordinate length of the
F-string is constant, as required. As we showed in Section
\ref{sec:Dbranes}, we expect cusps to form on a significant fraction
of such strings, in contrast to standard cosmic strings which can form
cusps only on closed strings.

%%%%%%%%%%%%%%%%%%%%%%%%%%%%%%%%%%%%%%%%%%%%%%%%%%%%%%%%%%%%%%%%%%%%%%%%%%%

\section{Phenomenology}

At the end of brane inflation a network of D- and F- strings is
expected to form. Although the intercommutation probability for two
D-stings is less than one~\cite{jjp05}, in the weak string coupling
limit $(g_s\rightarrow 0)$ it is always possible for an F- and D-
string to intercommute at a collision~\cite{Copeland:2007nv}, forming
a three string junction. As we have seen an F-string ending on two
such three string junctions will generically have cusps, at least in
the regime where the separation between the two D-strings is small,
i.e. when the distance between the two three string junctions is
small. We have seen that in this limit, the heavy D/FD-string behaves
as a single infinitely long string that is unaffected by the dynamics
of the F-string. At the level of approximation used here (of order
$g_s$), there is no energetic difference between the D-string and the
bound FD-string, however simulations~\cite{Sakellariadou:2008ay} have
shown that the FD-string will grow at the expense of the D string.
Explicit calculations of the intercommutation probability of F-strings
with multiple bound state strings has been done~\cite{jjp05} and in
the $g_s\rightarrow 0$ limit they are identical. Thus, an F-string can
intercommute with an FD-string to form a multiple bound state, with,
to order $g_s$, the same probability of intercommutation as with a
D-string. In effect then, we have two types of strings evolving in
such a network, heavy strings which are D or FD strings or multiple
bound states, and light F strings, which move in the background of the
heavy strings, but do not affect their dynamics.

For cusps to be significant to the dynamics of the F-string, we
require that the typical separation of the heavy strings is small
compared to the length of F-string stretched between them. This
condition is met early in the evolution of the string network~\cite{dvalietal},
where the typical inter-string distance is of the order of the symmetry
breaking scale~\cite{Vilenkin_shellard} and the string network is in
the friction dominated regime. As the heavy strings move apart, the F-strings 
ending on them will stretch, increasing the inter-string distance and reducing
the importance of cusps. This is in addition to the fact that as the
heavy strings move beyond the horizon scale, any effects of cusps would
be lost. Thus, phenomenological consequences of cusps from junctions on
cosmic superstrings will be most significant at early times, close to
the end of brane inflation. This is similar to the situation in cosmic strings 
when particle production from cusps on string loops is dominant in the friction
dominated regime and was used as a baryogenesis mechanism 
in Ref.~\cite{Brandenberger1991}.

The radiative properties of the cusps from F-strings are similar to
those of standard cosmic strings. In particular, the mechanism for the
emission of gravitational and particle radiation from cusps on loops
of cosmic
string~\cite{Damour:2004kw,Siemens:2006vk,Siemens:2006yp,Brandenberger}
also applies to cusps on strings between junctions. As in the standard
case the energy (and entropy) released by cusps goes into excitation
of all available fields. In the cosmic string case these are just the
standard model fields, however here they could include the
dilaton~\cite{Firouzjahi:2007dp},
Ramond-Ramond~\cite{Damour:1996pv,Sakellariadou:2004wq} and moduli
fields and all other fields present in the low energy limit of string
theory.  In particular, it is possible that gravitinos and other
stable supersymmetry particles may be produced, in addition to
standard model particles. Since this particle production would
predominately occur at early times, cosmological events such as
baryogenesis and big bang nucleosynthesis should be sensitive to the
presence of such cusps. This is in addition to the fact that cusps on
such a network would contribute to the gravitational wave
background~\cite{Damour:2004kw,Siemens:2006vk,Siemens:2006yp}.

Under S-duality the r\^ole of F- and D-strings is reversed; however the
results derived here apply with the heavy strings now being the F-strings
and the bound FD-string states. In this case the cusps would be present on
the light D-strings ending on three string junctions, which allows a
connection to be made with the cosmic strings of supergravity that
coincide with D-strings in this limit~\cite{Binetruy2004}, enalbing
their properties to be studied in the supergravity limit~\cite{Brax2006}.

%%%%%%%%%%%%%%%%%%%%%%%%%%%%%%%%%%%%%%%%%%%%%%%%%%%%%%%%%%%%%%%%%%%%%%%%%%%

\section{Conclusions}
We have shown how to characterise classical solutions to low energy
effective string actions in an intuitive geometric manner. We have
shown how the boundary conditions for strings ending on D-branes can
lead to cusps, points with luminous velocity, in such solutions, in an
analogous manner to cusps of standard cosmic string loops. In
particular, we have demonstrated that cusps would be a generic feature
of an F-string ending on two (parallel and stationary) D-strings. We
have then considered general three string junctions that are possible
in DBI-actions and have shown that the boundary conditions of such a
junction can similarly be characterised and understood in a geometric
setting, for the case of an F- and D-string meeting an FD-string. We
find that this system exactly reproduces the situation of an F-string
ending on two D-strings and hence a pair of such junctions would
generically include cusps. We have shown that this remains true even
when the D-strings are moving.

The relevance of such a scenario is that networks of F- and D-strings
are expected to form at the end of brane inflation. Collisions of such
networks would lead to pairs of three string junctions, each of which
would then be expected to have cusps, opening up a new energy loss
mechanism for such networks. Our new feature is in addition to cusps on 
string loops. Importantly the formation and existence
of cusps is expected to be most significant early in the evolution of
the network. The fact that the radiation from such cusps should
include all available fields present in the low energy string theory,
makes it possible that signatures of their presence would appear in
baryogenesis or big bang nucleosynthesis.

The extreme nature of cusps means that they are possible targets for
observations of string networks and here we have shown that it is, in
principle, possible to distinguish between standard cosmic strings
formed during cosmological phase transitions and cosmic superstring,
relics of brane inflation. The observation of three string junctions
would provide strong evidence  for
string theory. In future it might be possible to observe the emission of
radiation from cusps and the distribution of such cusps. 

\ack It is a pleasure  to thank M. Green and J. Polchinski
for enlightening discussions.  The work of M.S. is partially supported
by the  European Union through  the Marie Curie Research  and Training
Network {\sl UniverseNet} (MRTN-CT-2006-035863). This work is supported
in part by STFC.

\end{document}